\documentclass[twoside]{article}
\usepackage{fleqn,espcrc2}
\usepackage{graphicx}
\usepackage{rotating}

\title{Microwave Electrodynamics of the Antiferromagnetic Superconductor GdBa$_2$Cu$%
_3$O$_{7-\delta }$.}
\author{Lucia V. Mercaldo\thanks{
Also at Dipartimento di Fisica, Universit\`{a} degli Studi di Salerno,
Baronissi, Salerno I-84081, Italy.}, Vladimir V. Talanov\thanks{
Present address: Neocera, Inc., 10000 Virginia Manor Road, Beltsville, Maryland 20705.}, and Steven M. Anlage\address{Center for Superconductivity Research, Department of Physics,\\
University of Maryland, College Park, MD 20742-4111}}

\begin{document}

\begin{abstract}
The temperature dependence of the microwave surface impedance and conductivity are used to study the pairing symmetry and properties of cuprate superconductors. However, the superconducting properties can be hidden by the effects of paramagnetism and antiferromagnetic long-range order in the cuprates. To address this issue we have investigated the microwave electrodynamics of GdBa$_2$Cu$_3$O$%
_{7-\delta }$, a rare-earth cuprate superconductor which shows long-range ordered antiferromagnetism below $T_N$=2.2 K, the Neel temperature of the Gd ion subsystem. We measured the temperature dependence of the surface resistance and surface reactance of $c$-axis oriented epitaxial thin films at 10.4, 14.7 and 17.9 GHz with the parallel plate resonator technique down to 1.4 K. Both the resistance and the reactance data show an unusual upturn at low temperature and the resistance presents a strong peak around $T_N$ mainly due to change in magnetic permeability.
\end{abstract}
\maketitle

The analysis of the temperature dependence of the microwave surface
impedance and conductivity is one of the acclaimed methods used to extract information on the
pairing symmetry and properties of high temperature cuprate superconductors.
However things get more complicated when the material also develops magnetic
correlations, due to the localized moments of the rare-earth elements (RE).
The effects of paramagnetism and antiferromagnetic long range order may hide
the behavior of the superconducting screening length, influencing
conclusions about the pairing symmetry, as has been suggested for the
electron-doped Nd$_{2-x}$Ce$_x$CuO$_4$ \cite{cooper}.

To address this issue we have focused on the electrodynamic properties of
GdBa$_2$Cu$_3$O$_{7-\delta }$ (GBCO), where the Gd$^{3+}$ ions carry
magnetic moments which align parallel to the {\it c-}axis and order
antiferromagnetically below $T_N\simeq 2.2$ K in the three crystallographic directions \cite{neutron}.

The samples we have investigated are pairs of identical {\it c}-axis
oriented GBCO epitaxial films, laser ablated on (100)-cut LaAlO$_3$ single
crystal substrates. The film thickness is 300 nm, the superconducting
critical temperature measured by AC susceptibility is 92.5 K and the
transition width is 0.3 K.

We measured the effective (due to the finite film thickness) surface
impedance $Z_{Seff}\left( T,\omega \right) =R_{Seff}\left( T,\omega \right)
+iX_{Seff}\left( T,\omega \right) =\sqrt{i\omega \mu \left( T,\omega \right)
/\sigma \left( T\right) }\coth \left[ t\sqrt{i\omega \mu \left( T,\omega
\right) \sigma \left( T\right) }\right] $ \cite{matik} of the GBCO thin films
from 1.4 K\ to $T_c$ with the parallel plate resonator (PPR) technique \cite{ppr,lucia} at three different resonance frequencies with rf magnetic 
field in the $ab$ plane. Here the first factor on the right hand side is the bulk surface impedance $Z_S$ and the second factor is the finite thickness
correction ($t$ is the film thickness), and $\mu $ and $\sigma $ are
respectively the complex magnetic permeability and complex conductivity.
The PPR resonance frequency $f(T)$ and quality factor $Q(T)$ data are first converted to changes in surface reactance and surface resistance \cite{lucia} and then to absolute values using 
$X_{Seff}(77K)=49$ m$\Omega $ and $%
R_{Seff}\left(77K\right) =0.48$ m$\Omega $ measured at 10\ GHz by the
variable spacing parallel plate resonator technique \cite{vlad}.

\begin{figure}[htb]
\includegraphics[width=17.5pc]{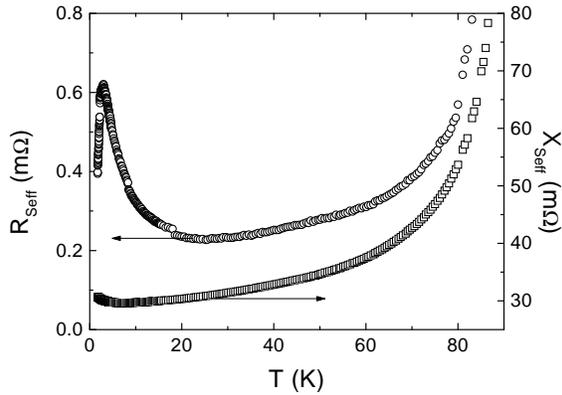}
\vspace{-4mm}
\caption{\label{fig:imped}
Effective surface resistance$R_{Seff}(T)$ and reactance $X_{Seff}(T)$ of GBCO at 10.4 GHz.}
\end{figure}

In Fig. \ref{fig:imped} we show $R_{Seff}(T)$ and $X_{Seff}(T)$ 
at 10.4 GHz over
the entire measurement temperature range. The high temperature behavior is
consistent with a $d$-wave temperature dependence for the surface impedance \cite{vlad}. The deviations from this behavior start below 30 K, where the
magnetic effects due to $\mu \left( T\right) $ come into play \cite{moss}.
Both $R_{Seff}(T)$ and $X_{Seff}(T)$ show a minimum at two different
temperatures, $T\simeq $ 25 K and $\simeq $ 7 K, respectively. Then $%
R_{Seff}(T)$ and $X_{Seff}(T)$ increase upon reducing the temperature, 
and a strong peak is observed in $R_{Seff}(T)$.

\begin{figure}[htb]
\vspace{3mm}
\includegraphics[width=18pc]{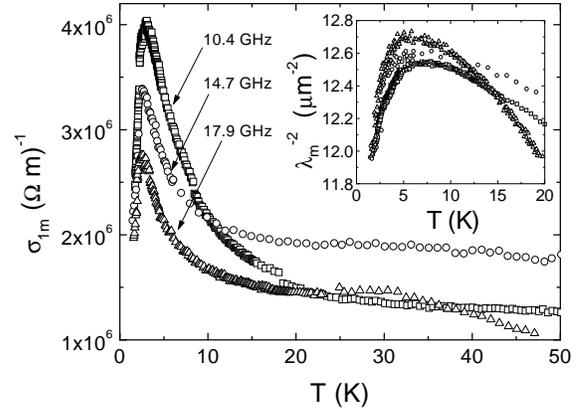}
\vspace{-6mm}
\caption{
Real part of the modified conductivity $\sigma_{1m}$ at different frequencies. Inset shows the rescaled imaginary part $\lambda_{m}^{-2}$ (same symbols as for the real part).}
\label{fig:cond}
\end{figure}

The same behavior is found at the other two frequencies with some extra
frequency dependence other than the trivial $X_S\sim \omega $ and $R_S\sim
\omega ^2$ observed for superconductors. This is clearly seen in Fig. \ref{fig:cond}, where we show the data at the three frequencies as modified complex conductivity $\sigma_m\left( T,\omega \right) $, defined through $Z_S\left( T,\omega \right)=\sqrt{i\omega \mu _0/\sigma _m\left( T,\omega \right) }$. The real part, $\sigma_{1m}=2R_S \omega \mu_0/X_S^3$, presents frequency dependent peaks around $T_N$. In the inset to Fig. \ref{fig:cond} 
we show a rescaled imaginary part $\lambda_{m}^{-2}=\sigma_{2m}\omega\mu_0=(\omega\mu_0/X_S)^2$, where the frequency dependence is less pronounced.

In conclusion strong unusual features are observed in the temperature dependence of surface impedance and conductivity for GBCO. The effects of paramagnetism and antiferromagnetism are shown to have a significant influence on $\lambda(T)$ and $R_S(T)$.

The authors want to acknowledge J. Claassen, M. Coffey, P. Fournier H. Harshevarden, M. Pambianchi, A. Pique, A. Porch and A. Schwartz.

\end{document}